\documentclass{elsart}
\usepackage{epsf}
\begin{document}
\begin{flushright}
KUPT 2000-02 \\
hep-ph/0009244\\
November 2000 
\end{flushright}
 
\begin{frontmatter}
\title{Analysis of the QCD-improved factorization in $B\rightarrow
J/\psi K$} 
\author{Junegone Chay\thanksref{KUPT}} and \author{Chul Kim} 
\address{Department of Physics,  Korea University, Seoul 136-701,
Korea}
\thanks[KUPT]{e-mail address: chay@kupt4.korea.ac.kr}
\begin{abstract}
We consider the exclusive decay $B\rightarrow J/\psi K$ using the
QCD-improved factorization method in the heavy quark limit. It is
shown that the decay amplitude is factorizable in this limit and
nonfactorizable contributions are calculable from first principles in 
perturbation theory. Also the spectator contributions at order
$\alpha_s$ are finite and suppressed in the heavy quark limit. We 
present the result at next-to-leading order in strong interaction, and
leading order in $1/m_b$ in the heavy quark limit. 
\end{abstract}
\end{frontmatter}

Exclusive nonleptonic decays of $B$ mesons have received a lot of
attention since they are observed in experiments at CLEO, BABAR and
BELLE \cite{exp}. They offer the opportunity to test the unitarity of
the Cabibbo-Kobayashi-Maskawa (CKM) triangle, the CP violation and to
observe new physics effects. However, quantitative understanding of
nonleptonic decays of $B$ mesons is difficult since there are final-state
interactions which are intrinsically nonperturbative in
nature. But when a $B$ meson decays into two light mesons, there is large
momentum transfer to final-state mesons. In this case, we can
systematically calculate the decay rates using perturbation theory in
the heavy quark limit.   

Beneke et al. \cite{beneke} considered general two-body
nonleptonic decays of $B$ mesons extensively including a light-light
meson system as well as a heavy-light system in the final state. The
general idea is that in the limit $m_b \gg \Lambda_{\mathrm{QCD}}$,
the hadronic matrix elements can be schematically represented as
\begin{eqnarray}
\langle M_1 M_2 |O|\overline{B}\rangle &=& \langle
M_1|j_1|\overline{B} \rangle \langle M_2|j_2 |0\rangle \nonumber \\
&\times& \Bigl[ 1 + \sum r_n \alpha_s^n +O(\Lambda_{\mathrm{QCD}}/m_b)
\Bigr],
\end{eqnarray}
where $M_1$, $M_2$ are final-state mesons and $O$ is a local
current-current operator in the weak effective Hamiltonian. If we
neglect radiative corrections in $\alpha_s$ and power corrections in
$\Lambda_{\mathrm{QCD}}$, we get the factorized result with a form
factor times a decay constant. At higher order in $\alpha_s$, this
simple factorization is broken, but the corrections can be calculated
systematically in terms of short-distance Wilson coefficients and
meson light-cone distribution amplitudes. We call this the
QCD-improved factorization.

When we use light-cone meson wave functions for exclusive decays,
the transition amplitude of an operator $O_i$ in the weak effective
Hamiltonian is given by
\begin{eqnarray}
\langle J/\psi K | O_i|\overline{B}\rangle &=& \sum_j
F_j^{B\rightarrow K} (m_{\psi}^2) \int_0^1 dx T_{ij}^I (x) \phi_{\psi}
(x) \nonumber \\
&&+ \int_0^1 d\xi dx du T_i^{II} (\xi, x,u) \phi_B (\xi) \phi_{\psi}
(x) \phi_K (u),
\label{imqcd}
\end{eqnarray}
where $F_j^{B\rightarrow K} (m_{\psi}^2)$ are the form factors for
$\overline{B}\rightarrow K$, and $\phi_M (x)$ is the
light-cone wave function for the meson $M$. $T_{ij}^I (x)$ and
$T_i^{II} (\xi, x,u)$ are hard-scattering amplitudes, which are
perturbatively calculable. The second term in Eq.~(\ref{imqcd})
represents spectator contributions.

This method works well for the case with two light mesons like
$\pi\pi$ or $\pi K$ \cite{beneke,chin}, in which the final-state
mesons carry large momenta. Interestingly enough, when there is a heavy
quark in the final state such as $\overline{B}\rightarrow D^+ \pi^-$,
this method still works when a spectator quark of the $B$ 
meson is absorbed by, say, a $D$ meson \cite{beneke,politzer}. In
Ref.~\cite{politzer}, they calculated the ratio $\Gamma
(\overline{B}\rightarrow D^{+} \pi^-)/\Gamma
(\overline{B}\rightarrow D^{+*} \pi^-)$ or $\Gamma
(\overline{B}\rightarrow D^{+(*)} \rho^-)/\Gamma
(\overline{B}\rightarrow D^{+(*)} \pi^-)$ since the Wilson
coefficients were known at leading order at the time. The absolute
branching ratios for $\overline{B} \rightarrow D^{+(*)} \pi^-$ were
calculated recently since the Wilson coefficients are known at
next-to-leading order accuracy \cite{beneke,chay}.\footnote{In
Ref.\cite{beneke}, Beneke et al. correctly pointed out the typos in
Ref.\cite{chay}.} However, when the spectator quark is absorbed by a
light quark, say, in $\overline{B} \rightarrow D \pi^0$,
nonfactorizable contributions are infrared divergent, and the
factorization breaks down.

When we consider the decay $\overline{B} \rightarrow J/\psi
K$, it looks ambiguous at first sight whether we can apply the same
method used in $\overline{B} \rightarrow \pi \pi$, or
$\overline{B}\rightarrow D^+ \pi^-$, since the
spectator quark in the $B$ meson goes into a light $K$
meson. However, what is special about $J/\psi$ is that the size of
the charmonium is so small ($\sim 1/\alpha_s 
m_{\psi}$) that the charmonium has a negligible overlap with the
($\overline{B}$, $K$)  system, hence enabling the same improved
factorization method in the decay $\overline{B}\rightarrow J/\psi
K$. As the explicit calculation shows below, the nonfactorizable 
contribution is infrared safe and the spectator contribution is
suppressed in the heavy quark limit. These facts indirectly justify
the use of the improved factorization formula Eq.~(\ref{imqcd}) in
$\overline{B} \rightarrow J/\psi K$. 

When the mass of the $J/\psi$ meson is not negligible, the light-cone
wave function of the $J/\psi$ meson should include higher-twist
contributions. The light-cone wave functions are obtained in powers of
$m_{\psi}/E$ or $\Lambda_{\mathrm{QCD}}/E$ where $E$ ($\sim m_b$) is
the energy of the $J/\psi$ meson. For $B$ decays into two light
mesons, the higher-twist contributions are negligible since they are
of order $\Lambda_{\mathrm{QCD}}/E$. However, for
$\overline{B}\rightarrow J/\psi K$, higher-twist contributions are
important. Therefore we expect that the decay rate using only the leading,
asymptotic wave function of $J/\psi$ will be smaller than the
experimental result.

Due to the nonzero mass of $J/\psi$, we can think of several
approximations with different limits. For example, we can take the
infinite mass limit of the $b$ quark in which $m_b$ goes to infinity
while $m_{\psi}$ is fixed ($m_{\psi}/m_b \rightarrow 0$). In this case
the result simply reduces to the case of $B\rightarrow \pi \pi$. But
this limit is hardly realized in nature and
it is not reasonable to compare the theoretical prediction in this
limit with experimental data. We can consider another limit in which 
$m_b$ goes to infinity, while $m_{\psi}/m_b$ is
held fixed. In this case, if $m_{\psi}$ is heavy enough, it
would be a better idea to employ a nonrelativistic wave function for
$J/\psi$. However, we can still use the light-cone wave function for
$J/\psi$. But we have to include the effects of higher-twist wave
functions in order to compare with experimental data reasonably since
these effects can be appreciable. 

From now on, we assume that, in the limit in which $m_b$ goes to
infinity, $m_{\psi}$ is heavy enough to regard the size of the
$J/\psi$ meson as small, but light enough to employ the leading-twist
light-cone wave function for $J/\psi$. It is difficult to satisfy this
interesting, but hypothetical limit in reality, but we will not go
further into the detail about this point. In this paper, we
present the decay amplitude at next-to-leading order in the strong
interaction, neglecting $O(\Lambda_{\mathrm{QCD}}/m_b)$ corrections,
employing the same technique to $\overline{B} \rightarrow J/\psi K$ as
in $\overline{B} \rightarrow \pi \pi$.  

One technical point is that we will keep the
mass $m_{\psi}$ of the $J/\psi$ meson in the hard scattering
amplitude.  In order for this to be theoretically consistent, we
also have to consider the higher-twist wave functions for
$J/\psi$ since higher-twist wave functions contain corrections of
order $m_{\psi}/m_b$. However, the main issue is whether the
hard-scattering amplitudes in the heavy quark limit are infrared
finite at each order in $\alpha_s$ and $1/m_b$. The effects of the
higher-twist wave functions correspond to the expansion in $1/m_b$,
and each wave function is, in general, an independent
function. Therefore we have to verify that the hard-scattering
amplitude convoluted with each independent wave function is
finite. What we pursue in this paper is whether the hard-scattering
amplitude convoluted with the leading-twist wave function for $J/\psi$
can be reliably calculated in the heavy quark limit. 
The inclusion of higher-twist wave functions to compare
with experimental data will be considered elsewhere.

The effective Hamiltonian $H_{\mathrm{eff}}$ for $\overline{B}
\rightarrow J/\psi K$ is written as 
\begin{equation}
H_{\mathrm{eff}} = \frac{G_F}{\sqrt{2}} \Bigl( V_{cb} V_{cs}^* (C_1
O_1 +C_2 O_2 ) -V_{tb} V_{ts}^* \sum_{i=3}^6 C_i O_i \Bigr),
\end{equation}
where $C_i$ are the Wilson coefficients at next-to-leading order
evaluated at the renormalization scale $\mu$. The relevant operators
in $H_{\mathrm{eff}}$ are given by
\begin{eqnarray}
O_1 &=& \overline{s}_{\alpha} \gamma^{\mu} (1-\gamma_5) c_{\alpha}
\cdot \overline{c}_{\beta} \gamma_{\mu} (1-\gamma_5) b_{\beta},
\nonumber \\
O_2 &=& \overline{s}_{\alpha} \gamma^{\mu} (1-\gamma_5) c_{\beta}
\cdot \overline{c}_{\beta} \gamma_{\mu} (1-\gamma_5) b_{\alpha},
\nonumber \\
O_3 &=& \overline{s}_{\alpha} \gamma^{\mu} (1-\gamma_5) b_{\alpha}
\cdot \sum_q \overline{q}_{\beta} \gamma^{\mu} (1-\gamma_5) q_{\beta},
\nonumber \\  
O_4 &=& \overline{s}_{\alpha} \gamma^{\mu} (1-\gamma_5) b_{\beta}
\cdot \sum_q \overline{q}_{\beta} \gamma^{\mu} (1-\gamma_5) q_{\alpha},
\nonumber \\  
O_5 &=& \overline{s}_{\alpha} \gamma^{\mu} (1-\gamma_5) b_{\alpha}
\cdot \sum_q \overline{q}_{\beta} \gamma^{\mu} (1+\gamma_5) q_{\beta},
\nonumber \\  
O_6 &=& \overline{s}_{\alpha} \gamma^{\mu} (1-\gamma_5) b_{\beta}
\cdot \sum_q \overline{q}_{\beta} \gamma^{\mu} (1+\gamma_5)
q_{\alpha},
\end{eqnarray}
where $\alpha$, $\beta$ are the color indices and the sum over $q$
runs over $u$, $d$, $s$, $c$ and $b$.

In calculating the
decay amplitude, we introduce the vector and the tensor decay
constants as \cite{ball1} 
\begin{eqnarray}
\langle J/\psi | \overline{c} (0) \gamma_{\mu} c(0) |0\rangle &=& -i 
f_{\psi} m_{\psi}\epsilon^*_{\mu},
\label{vector}\\
\langle J/\psi | \overline{c} (0) \gamma_{\mu}\gamma_{\nu} c(0)
|0\rangle &=& i f_{\psi}^T (\epsilon^*_{\mu} p_{\nu} -\epsilon^*_{\nu}
p_{\mu}), 
\label{tensor}
\end{eqnarray}
where $f_{\psi}$ is the decay constant which can be
determined from the leptonic decay of $J/\psi$. $f_{\psi}^T$ is the
tensor decay constant arising from the tensor current, which formally
depends on the renormalization scale. 

The distribution amplitudes on the light cone at leading-twist
accuracy are expressed compactly as \cite{ball2,benfeld}
\begin{eqnarray}
\langle J/\psi (p,\epsilon)| \overline{c}_{\alpha} (y) c_{\beta} (z)
|0\rangle &=& -\frac{i}{4} \int_0^1 dx e^{i(x p\cdot y + (1-x) p \cdot
z)} \nonumber \\
&&\times \Bigl[ f_{\psi} m_{\psi} \bigl( \FMslash{\epsilon}^*
\bigr)_{\beta \alpha} \phi_{\psi} (x) + f_{\psi}^T \bigl
( \FMslash{\epsilon}^* \FMslash{p} \bigr)_{\beta \alpha} \phi_{\psi}^T
(x) \Bigr]. 
\label{lcone}
\end{eqnarray}
In Eq.~(\ref{lcone}), $x$
is the momentum fraction of a $c$ quark inside the $J/\psi$ meson, and
the asymptotic wave functions $\phi_{\psi} (x)$, $\phi_{\psi}^T (x)$
for the $J/\psi$ meson are symmetric functions under $x\leftrightarrow
1-x$. The asymptotic form of the distribution amplitudes $\phi_{\psi}
(x)$ and $\phi_{\psi}^T (x)$ is the same, given as $\phi_{\psi} (x) =
\phi_{\psi}^T (x) = 6x (1-x)$. In the numerical analysis, we also
consider the wave function of the form $\phi_{\psi} (x) =\phi_{\psi}^T
(x) =\delta (x-\half)$, which appeals to the intuitive expectation of the
wave function.

If we neglect the radiative corrections and the power corrections in
$\Lambda_{\mathrm{QCD}}/m_b$, the factorized amplitude at leading
order is written as 
\begin{eqnarray}
iM_0 &=& -2i  f_{\psi} m_{\psi} \epsilon^*\cdot p_B
F_1 (m_{\psi}^2) \frac{G_F}{\sqrt{2}}\nonumber \\
&&\times \Bigl( V_{cb} V_{cs}^* (C_2 +\frac{C_1}{N} ) -V_{tb} V_{ts}^*
(C_3 + \frac{C_4}{N} +C_5 +\frac{C_6}{N}) \Bigr).
\label{tree}
\end{eqnarray}
We do not include the effects of the electroweak penguin operators
since they are numerically small. Here $N$ is the number of colors,
$m_{\psi}$ is the mass of $J/\psi$, 
and $\epsilon^{\mu}$ is the polarization vector of $J/\psi$. $F_i (q^2)$
$(i=0,1)$ are the form factors for $\overline{B}  \rightarrow K$,
which are given as 
\begin{eqnarray}
V_{\mu} &=&\langle K(p^{\prime}) | \overline{s} \gamma_{\mu} b|
B(p_B)\rangle \nonumber \\
&=&  
\Bigl[ (p_B +p^{\prime})_{\mu} -\frac{m_B^2-m_K^2}{p^2} p_{\mu} \Bigr]
F_1 (p^2) + \frac{m_B^2-m_K^2}{p^2} p_{\mu} F_0 (p^2),  
\label{vmu}
\end{eqnarray}
where $p= p_B -p^{\prime}$ is the momentum of $J/\psi$ with $p^2 =
m_{\psi}^2$.  From now on, we will neglect the kaon mass for simplicity. 

As we can see easily in Eq.~(\ref{tree}), this amplitude depends on
the renormalization scale $\mu$ since the Wilson coefficients depend
on $\mu$ while the matrix elements of the operators are replaced by
decay constants and form factors which are independent of
$\mu$. Therefore this amplitude is unphysical. However, if we include
the $\alpha_s$ correction to the amplitudes, it turns out that the
$\mu$ dependence of the Wilson coefficients is cancelled and the
overall amplitude is insensitive to the renormalization scale.

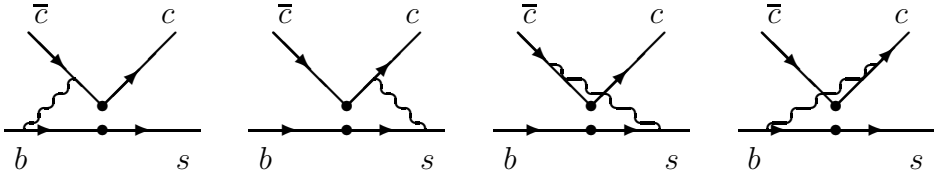
\begin{figure}
\begin{picture}(200,70)(-15,5)
\unitlength=0.65mm
\thicklines
\put(0,20){\line(1,0){40}}
\put(0,20){\vector(1,0){10}}
\put(20,20){\vector(1,0){10}}
\put(50,20){\line(1,0){40}}
\put(50,20){\vector(1,0){10}}
\put(70,20){\vector(1,0){10}}
\put(100,20){\line(1,0){40}}
\put(100,20){\vector(1,0){10}}
\put(120,20){\vector(1,0){10}}
\put(150,20){\line(1,0){40}}
\put(150,20){\vector(1,0){10}}
\put(170,20){\vector(1,0){10}}
\put(20,20){\circle*{2}}
\put(70,20){\circle*{2}}
\put(120,20){\circle*{2}}
\put(170,20){\circle*{2}}

\put(5,40){\line(1,-1){15}}
\put(5,40){\vector(1,-1){7.5}}
\put(20,25){\line(1,1){15}}
\put(20,25){\vector(1,1){7.5}}
\put(20,25){\circle*{2}}

\put(55,40){\line(1,-1){15}}
\put(55,40){\vector(1,-1){7.5}}
\put(70,25){\line(1,1){15}}
\put(70,25){\vector(1,1){9}}
\put(70,25){\circle*{2}}

\put(105,40){\line(1,-1){15}}
\put(105,40){\vector(1,-1){6}}
\put(120,25){\line(1,1){15}}
\put(120,25){\vector(1,1){7.5}}
\put(120,25){\circle*{2}}

\put(155,40){\line(1,-1){15}}
\put(155,40){\vector(1,-1){7.5}}
\put(170,25){\line(1,1){15}}
\put(170,25){\vector(1,1){12}}
\put(170,25){\circle*{2}}

\multiput(14.5,29)(-3,-3){4}{\oval(3,3)[tl]}
\multiput(11.5,29)(-3,-3){3}{\oval(3,3)[br]}

\multiput(75.5,29)(3,-3){4}{\oval(3,3)[tr]}
\multiput(78.5,29)(3,-3){3}{\oval(3,3)[bl]}
\multiput(111.5,32)(5,-3){5}{\oval(5,3)[tr]}
\multiput(116.5,32)(5,-3){4}{\oval(5,3)[bl]}
\multiput(178.5,32)(-5,-3){5}{\oval(5,3)[tl]}
\multiput(173.5,32)(-5,-3){4}{\oval(5,3)[br]}

\put(2,12){$b$}
\put(35,12){$s$}
\put(6,42){$\overline{c}$}
\put(32,42){$c$}

\put(52,12){$b$}
\put(85,12){$s$}
\put(56,42){$\overline{c}$}
\put(82,42){$c$}

\put(102,12){$b$}
\put(135,12){$s$}
\put(106,42){$\overline{c}$}
\put(132,42){$c$}

\put(152,12){$b$}
\put(185,12){$s$}
\put(156,42){$\overline{c}$}
\put(182,42){$c$}
\end{picture}
\caption{Feynman diagrams for nonfactorizable contribution at order
$\alpha_s$ in $\overline{B}\rightarrow J/\psi K$. The dots represents
the operators $O_i$ ($i=1,4,6$).}
\label{fig1}
\end{figure}

Nonfactorizable contributions at order $\alpha_s$ come from the
radiative corrections of the operators $O_1$, $O_4$ and $O_6$, and
the relevant Feynman diagrams are shown in Fig.~\ref{fig1}. The
radiative corrections with a fermion loop do not contribute due to the
color structure.  For each operator $O_1$, $O_4$ and $O_6$, if
we add all the diagrams in Fig.~\ref{fig1} and symmetrize the result
with respect to $x\leftrightarrow 1-x$, the infrared divergence of
each diagram cancels and the remaining amplitude is infrared
finite. One thing to note is that there appear
imaginary parts in the nonfactorizable contribution, which are due to
the final-state interaction. The strong phase can be calculated in the
QCD-improved factorization and it is important in exploring the CP
violation in nonleptonic decays.  

The decay amplitudes, in general, contain two terms
proportional to $F_0$ and $F_1$, which come from the matrix elements
for $\overline{B} \rightarrow K$. However, if we use the equation of
motion, we can find a relation between these terms, so that we can
write the decay amplitude which depends only on $F_0$, or $F_1$. In
the framework of the QCD-improved factorization with light-cone
distribution functions for the mesons, we assume 
that all the quarks inside a meson are on their mass shell. Therefore
we can safely use the equation of motion. For example, if we contract
$p_B^{\mu}$ to the current $\overline{s} \gamma_{\mu} (1-\gamma_5) b$,
we get 
\begin{equation}
p_B^{\mu} \overline{s}\gamma_{\mu}(1-\gamma_5) b = \overline{s}
\FMslash{p}_B (1-\gamma_5) b = m_b \overline{s} (1+\gamma_5) b.
\end{equation}
Similarly, if we contract the kaon momentum $p^{\prime\mu}$ to the
same current, we obtain 
\begin{equation}
\overline{s} \FMslash{p}^{\prime} (1-\gamma_5) b =0.
\label{vcon}
\end{equation}
Here we neglect the mass of the strange quark, and accordingly the
kaon mass, for
simplicity. When we combine Eqs.~(\ref{vmu}) and (\ref{vcon}), we
obtain the relation between the two form factors $F_0$ and $F_1$ as
\begin{equation}
p^{\prime} \cdot V = p_B \cdot p^{\prime} \Bigl[
\Bigl(1-\frac{m_B^2}{p^2} \Bigr) F_1 (p^2) +\frac{m_B^2}{p^2} F_0 (p^2)
\Bigr] =0.
\label{fcon}
\end{equation}
From this relation, the ratio $F_0/F_1$ is given by
\begin{equation}
\frac{F_0 (p^2)}{F_1 (p^2)} = 1-\frac{p^2}{m_B^2}.
\end{equation} 
This relation was also observed in calculating heavy-to-light form
factors in the large energy limit \cite{charles}. In parameterizing
these form factors using the QCD sum rule, this 
ratio is also valid to first order in $p^2$ \cite{ball3}. Note that,
as $p^2\rightarrow 0$, this ratio becomes 1 as it should be to remove
the pole at $p^2=0$.   

When we calculate the decay amplitude, some of the terms have the form
\begin{eqnarray}
&&(m_B^2 -m_{\psi}^2) \overline{s} \FMslash{\epsilon}^* (1-\gamma_5) b -
2 m_B\epsilon^* \cdot p_B \overline{s} (1+\gamma_5) b \nonumber \\
&=& 2\Bigl[
p\cdot p^{\prime} \overline{s} \FMslash{\epsilon}^* (1-\gamma_5) b
-\epsilon^*  \cdot p_B \overline{s} \FMslash{p} (1-\gamma_5) b \Bigr]
\nonumber \\ 
&=& 2 ( p\cdot p^{\prime} \epsilon^{*\mu} -\epsilon^*\cdot p^{\prime}
p^{\mu} ) \overline{s} \gamma_{\mu} (1-\gamma_5) b \equiv 2a^{\mu}
\overline{s} \gamma_{\mu} (1-\gamma_5) b.
\label{amu}
\end{eqnarray}
In deriving Eq.~(\ref{amu}), we use the equation of motion with $p_B = 
p+p^{\prime}$, $\epsilon^*\cdot p=0$ and neglect the light quark
masses and the light meson masses. The vector $a^{\mu}$ satisfies the
relation 
\begin{equation}
a\cdot p^{\prime} =0, \ \ a\cdot p=a\cdot p_B= -\epsilon^* \cdot p^{\prime}
m_{\psi}^2.
\end{equation}
Therefore we obtain the relation
\begin{equation}
a\cdot V = - \epsilon^* \cdot p_B m_{\psi}^2  \Bigl[
\Bigl(1-\frac{m_B^2}{p^2} \Bigr) F_1 (p^2) +\frac{m_B^2}{p^2} F_0 (p^2)
\Bigr] =0,
\end{equation}
as a result of Eq.~(\ref{fcon}). With this identity, in calculating
the decay amplitude, those terms
proportional to $\epsilon^*\cdot p_B \overline{s} \gamma
(1+\gamma_5)b$ can be replaced by the terms proportional to
$\overline{s} \FMslash{\epsilon}^* (1-\gamma_5) b$ or vice versa.

The full amplitude for $\overline{B}\rightarrow J/\psi K$ is written as
\begin{equation}
iM = -2i f_{\psi} m_{\psi} \epsilon^* \cdot p_B \frac{G_F}{\sqrt{2}} 
\Bigl[  V_{cb} V_{cs}^* a_2
-V_{tb} V_{ts}^* (a_3 + a_5) \Bigr] F_1 (m_{\psi}^2),
\label{total}
\end{equation}
where the coefficients $a_i$ ($i=2,3,5$) in the NDR scheme are given as
\begin{eqnarray}
a_2 &=& C_2 +\frac{C_1}{N} +\frac{\alpha_s}{4\pi} \frac{C_F}{N} C_1
\Bigl( -18 +12\ln \frac{m_b}{\mu} + f_I + f_{II}
\Bigr), \nonumber \\
a_3 &=& C_3 +\frac{C_4}{N} +\frac{\alpha_s}{4\pi} \frac{C_F}{N} C_4
\Bigl( -18 +12 \ln \frac{m_b}{\mu} +f_I + f_{II}
\Bigr), \nonumber \\
a_5 &=& C_5 +\frac{C_6}{N} -\frac{\alpha_s}{4\pi} \frac{C_F}{N} C_6
\Bigl( -6 +12 \ln \frac{m_b}{\mu} +f_I + f_{II}
\Bigr).
\end{eqnarray}

The function $f_I$ is given by 
\begin{eqnarray}
f_I &=& \int_0^1 dx\  \phi_{\psi} (x) \Bigl[ \frac{3(1-2x)}{1-x} \ln x
-3i \pi +3 \ln (1-r^2)+ \frac{2r^2 (1-x)}{1-r^2 x} \nonumber \\
&&+ \Bigl( \frac{1-x}{(1-r^2 x)^2} -\frac{x}{\bigl(1-r^2
(1-x)\bigr)^2} \Bigr) r^4 x \ln r^2 x +\frac{r^4 x^2 \Bigl(\ln (1-r^2)
-i\pi\Bigr)}{\bigl( 1-r^2 (1-x) \bigr)^2} \Bigr] \nonumber \\
&+&\frac{f_{\psi}^T m_c}{f_{\psi}m_{\psi}} \int_0^1 dx \
\phi_{\psi}^T (x) 4r^2 \Bigl[  \Bigl(
\frac{1}{1-r^2 (1-x)} -\frac{1}{1-r^2 x} \Bigr) \ln xr^2 \nonumber \\
&&-\frac{\ln (1-r^2)-i\pi}{1-r^2 (1-x)}  \Bigr], 
\end{eqnarray}
where $r=m_{\psi}/m_B$. The function $f_I$ is infrared
finite even when $r$ is nonzero. We organize the result such that it is
proportional to $F_1 (m_{\psi}^2)$, and all the terms involving $r$
go to zero as $r\rightarrow 0$. In this form, we can see clearly
that the coefficients $a_i$ in the limit $r\rightarrow 0$ is the same
as the coefficients for, say, $\overline{B} \rightarrow \pi
\pi$ \cite{beneke}. Note that the contribution from the tensor current
vanishes as $r\rightarrow 0$. This is trivial since the polarization
vector $\epsilon^{\mu}$ is proportional to the momentum $p^{\mu}$ in
the massless limit and the tensor current does not contribute at all
in this limit. 

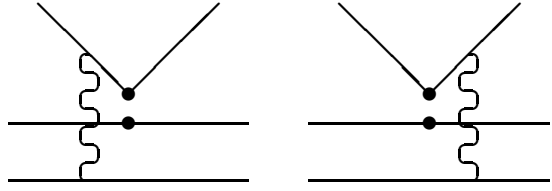
\begin{figure}
\begin{picture}(100,70)(-120,0)
\unitlength=0.8mm
\thicklines
\put(0,20){\line(1,0){40}}
\put(50,20){\line(1,0){40}}
\put(0,10.5){\line(1,0){40}}
\put(50,10.5){\line(1,0){40}}
\put(20,20){\circle*{2}}
\put(70,20){\circle*{2}}

\put(5,40){\line(1,-1){15}}
\put(20,25){\line(1,1){15}}
\put(20,25){\circle*{2}}

\put(55,40){\line(1,-1){15}}
\put(70,25){\line(1,1){15}}
\put(70,25){\circle*{2}}
\multiput(13.5,30)(0,-6){4}{\oval(3,3)[l]}
\multiput(13.5,27)(0,-6){3}{\oval(3,3)[r]}
\multiput(76.5,30)(0,-6){4}{\oval(3,3)[r]}
\multiput(76.5,27)(0,-6){3}{\oval(3,3)[l]}
\end{picture}
\caption{Feynman diagrams for the spectator contribution.}
\label{fig2}
\end{figure}

$f_{II}$ is obtained from the the spectator contribution to
$\overline{B}\rightarrow J/\psi K$, which is shown in
Fig.~\ref{fig2}. This corresponds to the second term in
Eq.~(\ref{imqcd}). When we symmetrize the amplitude with respect to
$x\leftrightarrow 1-x$, it turns out that there is no infrared
divergent part and the result is written as  
\begin{eqnarray}
iM_{II} &=& -2i f_{\psi} m_{\psi} \epsilon^* \cdot p_B
\frac{G_F}{\sqrt{2}} \Bigl( V_{cb} V_{cs}^* C_1 - V_{tb} V_{ts}^* (C_4
- C_6) \Bigr) \nonumber \\
&& \times \pi \alpha_s \frac{C_F}{N^2} \frac{f_B
f_K}{m_B^2} \frac{1}{1-r^2} \int d\xi \frac{\phi_B (\xi)}{\xi}  \int
du \frac{\phi_K(u)}{u} \int dx \frac{\phi_{\psi} (x)}{x},
\end{eqnarray}
where $\phi_B$, $\phi_K$ are the light-cone wave functions for the $B$
meson and the $K$ meson respectively. The spectator contribution
depends on the wave function $\phi_B$ through the integral
\begin{equation}
\int_0^1 d\xi \frac{\phi_B (\xi)}{\xi} \equiv \frac{m_B}{\lambda_B}.
\end{equation}
Since $\phi_B (\xi)$ is appreciable only for $\xi$ of order
$\Lambda_{\mathrm{QCD}}/m_B$, $\lambda_B$ is of order
$\Lambda_{\mathrm{QCD}}$. Therefore $f_{II}$ is given by
\begin{equation}
f_{II} = \frac{4\pi^2}{N} \frac{f_K f_B}{F_1 (m_{\psi}^2) m_B^2}
\frac{1}{1-r^2} 
\int_0^1 d\xi \frac{\phi_B (\xi)}{\xi} \int_0^1 dx
\frac{\phi_{\psi}(x)}{x} \int_0^1 du \frac{\phi_K (u)}{u}.
\end{equation}

The proof that the QCD-improved factorization method works for
$\overline{B}\rightarrow J/\psi K$ is sophisticated. There are two
independent parts which depend on  $f_{\psi} \phi_{\psi} (x)$ and
$f_{\psi}^T \phi_{\psi}^T (x)$. And each hard-scattering amplitude
amplitude proportional to $f_{\psi}$ or $f_{\psi}^T$ is infrared
finite. In both limits of 
$r\rightarrow 0$ and $r\neq 0$, the decay amplitudes are infrared
finite, justifying that the method of the QCD-improved
factorization can be applied in $\overline{B} \rightarrow J/\psi K$. 
This is the explicit proof of the QCD-improved factorization for  
$\overline{B} \rightarrow J/\psi K$ at leading-twist accuracy, which
was briefly mentioned in Ref.~\cite{beneke}.

One important ingredient in the above argument is that it holds true
only at leading-twist order. It means that we use only the
leading-twist wave functions, and higher-twist wave functions are
assumed to be suppressed. If $E$ is the energy of the $J/\psi$ meson,
it corresponds to neglecting those terms suppressed by $(m_{\psi},
\Lambda_{\mathrm{QCD}})/E \approx (m_{\psi},
\Lambda_{\mathrm{QCD}})/m_b$. In reality, we expect that higher-twist
effects are not negligible for appreciable values of $r$. 
When we consider higher-twist effects, we have to include the 
terms of order $\Lambda_{\mathrm{QCD}}/m_b$ in the higher-twist wave
functions as well as in the hard-scattering kernel. In order for the
QCD-improved factorization to work, the higher-twist effects should
also be infrared finite.

For numerical analysis, we use the following input parameters:
\begin{eqnarray}
m_b &=& 4.8 \ \mbox{GeV}, \ \ m_c = 1.5 \ \mbox{GeV}, \nonumber \\
m_B &=& 5.28 \ \mbox{GeV}, \ \ m_{\psi} =
3.1 \ \mbox{GeV}, \nonumber \\
f_{\psi} &=& 405 \ \mbox{MeV}, \ \ f_B = 180 \
\mbox{MeV}, \ \  
f_K =  160 \ \mbox{MeV}.
\end{eqnarray}
We also choose $\Lambda_{\overline{\mathrm{MS}}}^{(5)} = 225$ MeV, and
$\lambda_B \approx 300$ MeV.
For the form factors in $\overline{B} \rightarrow K$, we use the
values taken from Ref.~\cite{ball3}. 
\begin{equation}
F_1 (m_{\psi}^2) = 0.606, \ \ F_0 (m_{\psi}^2) = 0.418.
\end{equation}
And the CKM matrix elements are expressed in terms of the Wolfenstein
parameters \cite{wolfen} with $A=0.81\pm 0.06$ and $\lambda = \sin
\theta_C =0.2205\pm 0.0018$, and we fix them to their central values. 
For the wave function of $J/\psi$, we use employ two kinds of the wave
functions. One is the asymptotic function $\phi_{\psi} (x) =
\phi_{\psi}^T (x) =6x(1-x)$, and the other is $\phi_{\psi} (x) =
\phi_{\psi}^T (x) =\delta (x-1/2)$, which is more intuitive. 

In order to extract the magnitude of the tensor decay constant
$f_{\psi}^T$, we contract
Eq.~(\ref{tensor}) with $p^{\nu}$, and use Eq.~(\ref{vector}) and the
equation of motion. Then we get the relation
\begin{equation}
2m_c (\mu) f_{\psi} = m_{\psi} f_{\psi}^T (\mu),
\end{equation}
from which, we use
\begin{equation}
f_{\psi}^T m_c = \frac{2m_c^2 f_{\psi}}{m_{\psi}}.
\end{equation}
The numerical result of the coefficients $a_i$ is
summarized in Table~\ref{table1}. We can see that the coefficients
$a_i$ do not depend sensitively on the choice of the wave functions
since the numerical values with different choices of the wave
functions for $J/\psi$ are not much different.

\begin{table}
\caption{The coefficients $a_i^{1,0}$ at $\mu=m_b$ and $m_b/2$ with
different wave functions of $J/\psi$. The values in the second and the
third columns are the values with  $\phi_{\psi}(x) =6x(1-x)$ and the
fourth and the fifth columns are the values with the delta function.}
\vspace{7mm}
\begin{tabular}{ccccc} \hline    
$a_i$ & $\mu=m_b$ & $\mu=m_b/2$ & $\mu=m_b$, delta &
$\mu=m_b/2$, delta \\   \hline
$a_2\times 10^3$&72.4$-$60.0$i$&48.3$-$79.6$i$&77.3
$-$60.1$i$&54.9$-$79.8$i$ \\ 
$a_3 \times 10^3$& 4.01$+$1.39$i$&5.37$+$2.45$i$& 3.90
$+$1.39$i$&5.17$+$2.46$i$ \\ 
$a_5 \times
10^3$&$-$4.95$-$1.71$i$&$-$6.03$-$3.97$i$&$-$4.80$-$1.71$i$&
$-$5.70$-$3.98$i$ \\  \hline
\end{tabular}
\label{table1}
\vspace{5mm}
\end{table}

With these coefficients $a_i$, we can calculate the branching
ratio. The branching ratio from experiment is given by \cite{pdg}
\begin{equation}
{\mathrm{Br}} (\overline{B} \rightarrow J/\psi K) = (8.9 \pm 1.2) \times
10^{-4}.
\end{equation} 
Our result shows that, for $\phi_{\psi} (x) = 6x(1-x)$,
the branching ratio is $1.03 \times 10^{-4}$ for $\mu=m_b$ and
$1.05\times 10^{-4}$ for $\mu=m_b/2$, which are about eight times
smaller than the experimental result. Though the theoretical treatment
of the QCD-improved factorization in $\overline{B} \rightarrow J/\psi
K$ is improved, the theoretical result still does not saturate the
experimental result. This is expected since we do not include
higher-twist effects which are supposed to be rather appreciable. In
order to have a reasonable comparison, we need to include higher-twist
effects, which is going to be pursued elsewhere.

\begin{figure}[htb]
\begin{center}
\mbox{\epsfxsize 5in
\epsfbox{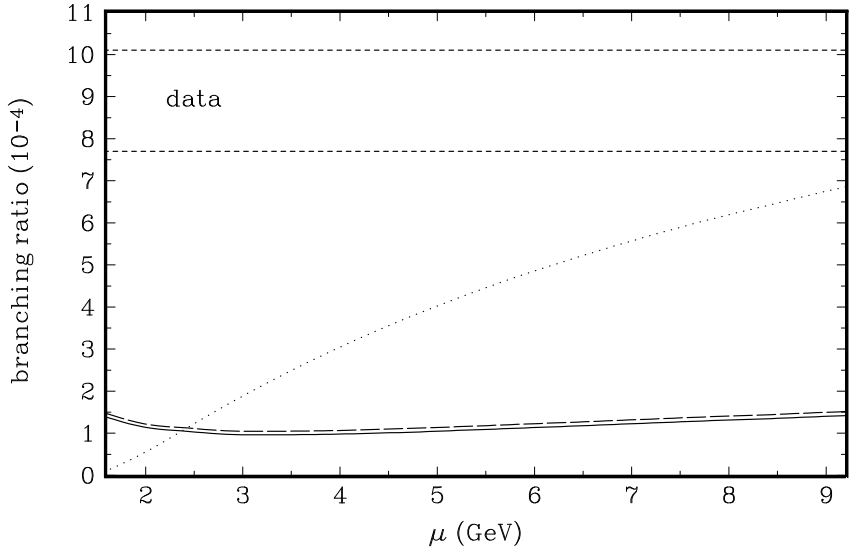}}
\end{center}
\vskip0.5cm
\caption{Dependence of the branching ratio on the renormalization
scale $\mu$. The solid line uses the wave function $\phi_{\psi}(x)=
6x(1-x)$, the long-dashed line uses $\phi_{\psi}(x) =\delta
(x-\half)$. The dotted line represents the result without the
$\alpha_s$ correction. The band with short dashes represents the
experimental data at $1\sigma$ level.}
\label{fig3}
\end{figure}

In Fig.~\ref{fig3}, the dependence of the branching ratio on the
renormalization scale is shown. The horizontal short-dashed lines in
Fig.~\ref{fig3} show the experimental branching ratio at $1\sigma$
level. We also show the factorized result without nonfactorizable
contribution. Of course, this quantity is unphysical and very
sensitive to the renormalization scale, as we can see clearly in
Fig.~\ref{fig3}. On the other hand, for the full branching ratio
calculated at next-to-leading order, the dependence of the branching
ratio on the renormalization scale is very mild.

Recent observation of CP asymmetry in $B\rightarrow J/\psi K_S$
\cite{cp} attracted some interests in the possibility of new physics
\cite{cpth}. In this scheme in the standard model, there is no
contribution to CP asymmetry in the decay amplitude since the CKM
matrix elements involved here are all real. The CP asymmetry totally
comes from $B-\overline{B}$ mixing, and we have not considered this
issue here. 

In conclusion, we have considered the effect of nonfactorizable
contributions in $\overline{B} \rightarrow J/\psi K$ in the heavy
quark limit. In this limit, nonfactorizable contribution can be
systematically calculated from first principles using perturbation
theory. It is shown that the decay amplitude is factorizable at
next-to-leading order in the strong interaction and at
leading order in $\Lambda_{\mathrm{QCD}}/m_b$. The nonfactorizable
contribution and the spectator contribution are infrared finite. 

In general, when the $B$ meson decays
into two mesons in which one is heavy, if the spectator quark of the
$B$ meson goes into the light meson, the decay amplitude is not
factorizable. The overlap of the heavy meson with the remainder is
sizable and the soft gluon exchange gives a significant
effect. However, when the heavy meson is the charmonium, the
factorization method still works at leading-twist order. This
physically means that the size of the charmonium is so small that the
overlap with other mesons is small. Therefore the QCD-improved
factorization method holds in $\overline{B} \rightarrow J/\psi K$,
in contrast to naive expectations.

In spite of the theoretical improvement mentioned so far, the
theoretical branching ratio is about eight times smaller than the 
experimental result. This is typical in class II decays in which there
is color suppression in the lowest order contribution. And the
$\alpha_s$ corrections give a significant change in magnitude and a
significant strong phase compared to the leading-order result
corresponding to naive factorization. 

Another reason why the theoretical result is small compared to the
experimental data is that we used only the leading-twist wave
functions. In a realistic case, higher-twist effects are not
negligible. Khodjamirian and R\"{u}ckl \cite{kr} considered
nonperturbative effects in $B\rightarrow J/\psi K$ using the QCD sum
rules, and they found that nonperturbative effects including the
higher-twist wave functions could indeed be large. Also in the
inclusive production of $J/\psi$, the naive factorization does not
explain the large production rate. It was suggested that the octet
contribution would enhance the rate \cite{pw,ben2}. This contribution
comes from the nonleading Fock state $|g \overline{c} c\rangle$ state, 
which constitutes higher-twist effects, and it may not be negligible
in the decay $\overline{B} \rightarrow J/\psi K$ either.  However, the
method of the QCD-improved factorization can be safely applied to
$\overline{B} \rightarrow J/\psi K$ at leading-twist order and  it is
remarkable that we obtain a very significant correction to naive
factorization.  

{\bf Note added}: While this paper has been written, Ref.~\cite{cheng}
appeared, in which the decay amplitude for $\overline{B} \rightarrow
J/\psi K$ are computed including chirally enhanced
contributions. Their result looks different from ours, but the
definition of $f_I$ and $g_I$ are rather ambiguous since they can be
switched back and forth using the equation of motion discussed in this
paper. Furthermore, they put $f_{\psi}^T m_c/f_{\psi} m_{\psi} =2x^2$ using
the equation of motion. That is why there is no $f_{\psi}^T$ in their
result. However, when all is taken into account, both results are the
same.
 
\section*{Acknowledgements}
The authors were supported in part by the Ministry of Education grants
KRF-99-042-D00034 D2002. C.K. is partially supported by the Korea
University Research Fund. The authors would like to thank Pyungwon Ko
and Hai-Yang Cheng for stimulating discussions.

\end{document}